# Sketching a Model on Fisheries Enforcement and Compliance

# A Survey

Manuel Coelho

coelho@iseg.ulisboa.pt

José António Filipe

jose.filipe@iscte-iul.pt

Manuel Alberto M. Ferreira

manuel.ferreira@iscte-iul.pt

## Abstract

Monitoring and enforcement considerations have been largely forgotten in the study of fishery management. This paper discusses this issue through a model formalization to show the impacts of costly, imperfect enforcement of law on the behavior of fishing firms and fisheries management. Theoretical analysis merges a standard bio-economic model of fisheries (Gordon-Schaefer) with Becker's theory of Crime and Punishment.

***Keywords*** - *Fisheries, Crime, Punishment, Enforcement, Becker's Theory.*

## 1. Introduction

Illegal fishing covers a wide range of behavior, which can take place at different levels: local, national, and international. Illegal fishing always existed, but, in recent decades, there has been a sharp rise in violating activities, due to several factors. Technical progress in motorization, freezing techniques, improved gear, new forms of stocks detection and information made it easier. But also, the new Law of the Sea (1982), generating a "creeping jurisdiction" process that seems to give an end to the principle of open access, is in the roots of this phenomenon.

Obviously, it's impossible to quantify or qualify infringements. They are known to take place at all levels and take different forms at different times; some violations are detected but many remain unnoticed. Infringements take the traditional forms of fishing over the quota or using non-permitted mesh size but are also in situations of non-permitted bycatches or transshipment, even in the fake world of convenience flags. The possibilities of fraud after landing are enormous.

Illegal fishing is a problem because it undermines efforts to conserve and manage fish stocks. Destruction of fishing-grounds seriously harms efforts to replenish stocks and diminishes social perspectives about economic returns, in both the short and long term.

The enforcement of Law is the other side of the mirror. The objective of monitoring and surveillance is to deter/detect infringements and to encourage compliance with the rules. The monitoring of fishing activities combines prevention, penalties, and the development of a sense of responsibility. Fisheries control regimes aim to contribute to resource management as a complement of other tools of the conservation policy, to discourage the inclination to infringe regulation, guaranteeing fair and transparent enforcement, to impose penalties on wrongdoers and, incidentally, also helps improve scientific knowledge.

In the context of Fisheries Economics, the monitoring problem can be seen as an externality arising when exclusive property rights are absent (Cheung, 1970). And that absence depends on, among other things, the costs of defining and enforcing exclusivity. Most of the literature on fisheries management and regulation implicitly assumes law can be perfectly and cost-less enforced.

Even if such costs and imperfections are recognized they are not usually incorporated in the analysis to show how agents' behavior and management policies are affected by their presence. Despite the astute observations in the traditional literature (Scott,1979), enforcement considerations have been largely ignored in Fisheries Economics.

This paper reviews the relevant literature and explores this issue with a formal model of fisheries law enforcement. Based on the mid-eighties analysis of Sutinen and Andersen, it merges the standard Gordon-Schaefer fisheries model with the insights of the so-called "Theory of Crime and Punishment" of Becker.

The paper has the following structure: Firstly, the model is introduced. The agent behavior in a situation of crime is discussed and a formulation of fisheries enforcement cost function is derived. Then follows the optimal management policy description when the enforcement costs are considered. Optimal control theory is introduced to solve the model and the results are discussed. Finally, the paper analyses the methods to improve compliance with fisheries regulation. Several policy prescriptions are made.

## 2. Illegal Behavior and Enforcement Costs

Despite the enormous volume of literature on Fisheries Economics, only a few papers are devoted to the issue of enforcement. According to Sutinen and Hennessey (1986), it has always been "the neglected element in fishery management".

The fundamental problem in fisheries management is to obviate the tendency towards overexploitation of the resources under open access. Starting at any initial stock size, a means must be found to reduce catch rates. Regulation methods used to curb this tendency of overfishing and overcapacity includes gear restrictions, area, and seasonal closures, TACs, ITQs, limiting entry and other forms of reducing fishing effort. Assume that whatever means are applied to reduce catch rates, any catch level above the level of the permitted quota for a certain fishing, $q^*$, is illegal.

If we suppose a system of individual non-transferable quotas, the amount of the individual firm catches above its quota $(q_i - q_i^*)$ is illegal.

If detected and convicted, a penalty fee is imposed on the firm in an amount given by f,

$f = f(q_i - q_i^*)$ where $f > 0$, if $q_i > q_i^*$ and $f = 0$, otherwise; and $\frac{\partial f}{\partial q} \geq 0$; $\frac{\partial^2 f}{\partial q^2} \geq 0, \forall q_i > q_i^*$ (2.1).

We assume that the function $f(\cdot)$ is continuous and differentiable for all $q_i > q_i^*$. This penalty fee has a finite upper bound and each firm is assumed to face the same penalty fee schedule.

An individual firm's profit before penalty is given by:

$$\Pi^i(q_i, x) = pq_i - c^i(q_i, x) \quad (2.2),$$

where *p* denotes the price of fish, x is the size of fish stock and *c(.)* is the cost function. We assume that firms are price takers.

In an imperfect law enforcement regime, not every violator is detected and convicted. Let the probability of detection and conviction be given by $\theta$, and, to simplify, let us assume that all firms face the same probability.

If detected and convicted of a violation, a firm's profit will be:

$$\Pi^i(q_i, x) - f(q_i - q_i^*) \quad (2.3);$$

if not,

$$\Pi^i(q_i, x) \quad (2.4).$$

So, expected profits are:

$$\theta \left[ \Pi^i(q_i, x) - f(q_i - q_i^*) \right] + (1 - \theta) \Pi^i(q_i, x) \quad (2.5).$$

Assuming firms are risk neutral and maximizing expected profits, each $q_i$ is determined by the first order condition (subscripts other than *i* denote partial derivatives),

$$\Pi_q^i(q_i, x) \geq \theta f_q(q_i, x) \quad (2.6).$$

The solution to (2.6) for one form of the marginal penalty schedule, $f_q$, lead to the result:

- If there were no penalty for fishing beyond legal quota, or if there were no probability of being detected and convicted (f = 0 or $\theta$ = 0) the firm would set its catch at the open access catch rate, $q_i^0$. For a given stock size (x), the firm sets its catch rate at a level more than its quota, where marginal profits equal the expected marginal penalty.

The first order condition (2.6) can be solved for a firm catch rate, as $q_i = q_i(\theta, x, q_i^*)$. The catch rate also depends on price, production cost parameters and parameters of the penalty fee schedule, but these are suppressed for notational simplicity. Note the following properties, important for the discussion, see Sutinen and Andersen (1985):

- An increase in the probability of detection and conviction decreases, or leaves unchanged, a firm catch rate, as the expected marginal penalty schedule becomes steeper.
- An increase in the stock size shifts up the marginal profit schedule and increases, or leaves unchanged, a firm's catch rate.
- An increase in the quota shifts the expected marginal penalty to the right and increases a firm's catch, so long as the initial catch rate is inferior to the bionomic equilibrium (open access situation).

Aggregating the catches for all firms in the fishery yields the aggregate catch function $q = q\,(\theta, x, q^*)$, where $q^* = \sum q_i^*$, for the $N$ firms operating in the fishing ground. Assuming there is a sufficient heterogeneity across firms to allow this equation to be continuous, therefore inverse forms exist and $\frac{\partial q}{\partial \theta} < 0, \frac{\partial q}{\partial x} > 0, \frac{\partial q}{\partial q^*} > 0$.

To detect and convict violators require inputs (aircraft, patrol vessels, police, and judicial personnel). Let the quantities of such inputs be represented by a vector $k$, which has an associated vector of unit prices $w$. The probability of detecting and convicting fraud is assumed to depend positively on the inputs.

Assuming the least cost combination of $k$ is chosen for each level of $\theta$, there is an enforcement cost function e($\theta$), where $\frac{\partial e}{\partial \theta} > 0, \frac{\partial^2 \theta}{\partial \theta^2} \geq 0$ and, using the inverse, enforcement costs can be represented by $E\,(q, x, q^*)$.

The following properties hold:

- A reduction in the catch level (below the open access level for a given size and quota) requires an increase in enforcement costs.
- Increase in the fish stock or quota requires greater enforcement costs to achieve a given catch level (note that this ignores some economies of scale in enforcement).
- The size of the quota also affects enforcement.

## 3. Optimal Policy

Now, we must investigate how optimal management policies are affected by costly, imperfect enforcement.

Optimal policies are based on the usual criterion of maximizing the discounted sum of net social benefits. In each period these net benefits are given by:

$$\int_0^q p(q)dq - c(q,x) - E(q,x) \quad (3.1),$$

where $p(q)$ is the inverse demand function, $c(q,x)$ is the aggregate cost catch function and $E(q,x)$ is the enforcement cost function.

The aggregate cost function depends on the fixed set of quotas and doesn't include penalty fees. These are excluded since they are transfers from fishing firms to general treasury. Since quota allocation is assumed exogenously determined $q^*$ is suppressed as argument in the enforcement cost function.

Now we introduce the usual *stock dynamics standard differential equation*, where $F(x)$ is the natural growth rate:

$$\dot{x} = F(x) - q \quad (3.2).$$

Optimal politics are found as a solution of the problem:

$$\text{Max} \int_0^\infty (\int_0^q p(q)dq - c(q,x) - E(q,x))e^{-\delta t} dt \quad (3.3),$$
$$\text{sub. to} \quad \dot{x} = F(x) - q$$

where $\delta$ is the interest rate.

The Lagrangian function is:

$$L(q,x,\lambda) = \int_0^\infty (\int_0^q p(q)dq - c(q,x) - E(q,x))e^{-\delta t} - c(q,x) - E(q,x) dt - \lambda(\dot{x} - F(x) + q) \quad (3.4)$$

or

$$L(q,x,\lambda) = \frac{\int_0^q p(q)dq - c(q,x) - E(q,x)}{\delta} - \lambda(\dot{x} - F(x) + q) \quad (3.5)$$

First order conditions are:

$$\begin{aligned} p - c_q - E_q - \lambda &= 0 \\ \lambda &= c_x + E_x + \lambda(\delta - F_x) \end{aligned} \quad (3.6),$$

being $\lambda$ the Lagrange multiplier (again, subscripts other than $i$ denote partial derivatives).

Solving (3.6), we are conducted to a *transformed golden rule*:

$$\delta - F_{x^{**}} = -\frac{c_{x^{**}} + E_{x^{**}}}{p^{**} - (c_{q^{**}} + E_{q^{**}})} \quad (3.7).$$

This determines the steady state optimal size stock, $x^{**}$, the optimal catch rate, $q^{**}$, and resulting price, $p^{**}$.

We can derive interesting conclusions if we compare this, with the situation where we assume costless and perfect enforcement, that is, when catch rates are perfectly controlled at zero cost. In this case the condition for optimality is the *usual modified golden rule,* see Clark and Munro (1975)*:*

$$\delta - F_{x^{***}} = -\frac{c_{x^{***}}}{p^{***} - c_{q^{***}}} \quad (3.8),$$

where *x*\*\*\* is the optimal stock size, *q*\*\*\* the optimal catch rate and *p*\*\*\* the resulting price.

By comparing the two golden rules it can be

concluded that:

- the presence of costly, imperfect enforcement results in a smaller optimal stock size than otherwise: $x^{**} < x^{***}$.
- similarly, higher enforcement costs result in a lower optimal stock.

The rationale is not difficult to follow.

If some kind of quota system is in effect to ration access, enforcement activity would involve monitoring compliance with these quotas and assigning penalties on those found in violation. If quotas were so large as being consistent with free access equilibrium, enforcement costs would be zero because no enforcement would be necessary to ensure compliance. But moving away from free access equilibrium increases both net benefits and enforcement costs. For this model, as the steady-state population size is increased, marginal enforcement costs increase, and marginal net benefits decrease. At the efficient population size, with enforcement costs, the net marginal benefit equals the marginal enforcement costs. This necessarily involves a smaller population size than the efficient population size ignoring enforcement costs, because the latter occurs when the marginal net benefit is zero (Tietenberg, 2003).

Sutinen and Andersen (1985) also compare the catch rates for costless, perfect enforcement with costly, imperfect enforcement. They conclude that depends on whether the stock sizes are above or below the so-called MSY (Maximum Sustainable Level). For the most usual case, that is, $x^{**}<x^{***}<x MSY$, the catch rate in the situation of costly enforcement costs is lower than the case without consideration of enforcement costs.

## 4. Enforcement and Compliance

Besides the intrinsic value of the model important research questions are suggested.

The enforcement issue points out another advantage of private property rights-based management: they are self-enforcing. This may represent an important step to proceed in the discussion of regulatory instruments. First, if enforcement costs are significant, the more common forms of regulation (command and control tools as TACs, mesh size or areas/seasons closures, for example) should require further re-evaluation.

Usually, they are detracted because they are not economically efficient. But is also commonly recognized that costs of enforcement are weaker in these cases. Second, the analysis of Individual Transferable Quotas reveals the equivalence between ITQs and taxes. But, with the consideration of enforcement costs, this may not hold. The analysis of Sutinen and Andersen, with non-transferable quotas, appears to parallel the case of taxes more closely. Investigation on the ITQs case is still a work in progress. In any case, the reduced costs of enforcement favored this tool. As the fishermen are given almost private property rights of resource use, this means that some kind of autoregulation is guaranteed. In theory, this engages fishermen in compliance with the regulation and diminishes enforcement costs. In practice, the implementation of ITQs systems is confronted with a lot of problems including illegal behavior (Copes, 1986).

In another dimension, this approach also reveals the importance of empirical studies trying to estimate the factors that ensure compliance with the regulation. These studies give important basis for public authorities' decision about the actions to be implemented.

Stigler (1970) argues that public authorities have four basic means to improve compliance:

- minimize the chances that violations will go undetected,
- maximize the probability that sanctions will follow the detection of violations,
- speed up the process from time to detection to assignment of sanction,
- make the sanctions large.

There is dispute among experts about the best alternatives. Some scholars have argued that the probability of being detected is more important than the size or magnitude of the sanction, while others argue that making the charging time follow as closely as possible to the detection of illegal behavior is the most important factor in enhancing compliance.

Others, also, put in evidence the level of expenditure oriented to monitoring activities.

Empirical evidence? The difficulties in getting data to this kind of research are very problematic. But we stress the necessity of more studies, see, for example, Sutinen and Hennessey (1986) or Sutinen and Gauvin (1989).

Econometric studies have demonstrated that all factors are significant. Especially the one that states fishermen perceived probability of detection and conviction affects their violation rate as predicted by the theory. The higher the probability, the lower the rate of violation. So, enforcement and other measures of control to increase perceived probabilities, enhance compliance.

Personal characteristics of fishermen also are expected to influence compliance behavior. In the study of Sutinen and Gauvin (about lobstermen in Massachusetts) the pattern is that fishermen who have more illegal landings are thought to be older and to fish fewer days. At the same time, they are thought to be in the fishery for the short term. This could be a portrait of a serious group of violators who are about to retire and are grasping at short time gains because they do not feel they will be around to reap the long-term benefits of conservation. But we can also expect that fishermen with more years in the fishery and more income dependent on a certain fishery have stronger conservation motives. So, studies must investigate these issues.

## 5. Final Remark

In the Portuguese case, after a significant process of modernization of the surveillance structures, several problems persist.

The European Commission gave the financial support to guarantee the indispensable means of surveillance and control and increased the deterrence capacity of control in member states (in a uniform way, which is important) and the transparency and trust between partners. That lead to the increase of the probability of detection to deter criminal behavior and increased compliance with regulation. But the difference in the judicial administration maintains.

In the Portuguese case, the dispersion of surveillance and control activities between several agencies (Maritime Authority, Ports Administration, different Policies' corps) is, always, referred as a fundamental root of inefficiency. The stakeholders put, also, in evidence, the delayed application of regulation by the tribunals.